# A Mathematical Model of Motion Sickness in 6DOF Motion and Its Application to Vehicle Passengers

T. Wada *†, Norimasa Kamiji‡, and Shun'ichi Doi‡

*† Ritsumeikan University*

*1-1-1 Noji-higashi, Kusatsu, Shiga, Japan*

*‡ Kagawa University*

*2217-20 Hayashi-cho, Takamatsu, Kagawa, Japan*

**Abstract**

A mathematical model of motion sickness incidence (MSI) is derived by integrating neurophysiological knowledge of the vestibular system to predict the severity of motion sickness of humans. Bos et al. proposed the successful mathematical model of motion sickness based on the neurophysiological mechanism based on the subject vertical conflict (SVC) theory. We expand this model to 6-DOF motion, including head rotation, by introducing the otolith-canal interaction. Then the model is applied to an analysis of passengers' comfort. It is known that the driver is less susceptible to motion sickness than are the passengers. In addition, it is known that the driver tilts his/her head toward the curve direction when curve driving, whereas the passengers' head movement is likely to occur in the opposite direction. Thus, the effect of the head tilt strategy on motion sickness was investigated by the proposed mathematical model. The head movements of drivers and passengers were measured in slalom driving. Then, the MSI of the drivers and that of the passengers predicted by the proposed model were compared. The results revealed that the head movement toward the centripetal direction has a significant effect in reducing the MSI in the sense of SVC theory.

## 1. Introduction

Carsickness obviously decreases the comfort of humans in a vehicle. It is necessary to clarify its mechanism and to develop a reduction method. The recent success of ride comfort has been realized by vibration analysis and subjective evaluation (Donohew & Griffin, 2004). Recently, research has studied driver's perceptual and cognitive characteristics to analyze ride comfort. For example, Muragishi et al. (2007) investigated the driver's sensitivity to vehicle motion by visual and motion inputs.

The prediction of motion sickness or ride comfort of vehicles is important to create comfortable vehicle motion. For estimating the severity of motion sickness, many studies revealed the motion sickness sensitivity by using the frequency and the magnitude of acceleration (Griffin, 1990). For example, O'Hanlan and McCauley (1974) obtained the motion sickness incidence (MSI), defined by the percentage of vomiting people by whole body vibration, from experiments in sinusoidal vertical motion.

The sensory rearrangement theory postulated that motion sickness is provoked by accumulation of the conflict between sensory information from the vestibular system and the estimated sensory information from an internal model (Reason, 1978). Bles et al. (1998) proposed the subjective vertical conflict (SVC) theory by employing errors between sensed and estimated vertical or gravitational directions as the conflict in the sensory rearrangement theory. Bos and Bles (1998) proposed the mathematical model to estimate the MSI in 1-DOF linear motion based on the SVC theory. To apply the results to analyzing the effect of head rotation including the car driver's or passenger's head, expansion to 6DOF motion in three-dimensional space is needed. Thus, Kamiji et al. (2007) proposed a 6DOF-SVC model that can estimate the MSI in 6-DOF motion.

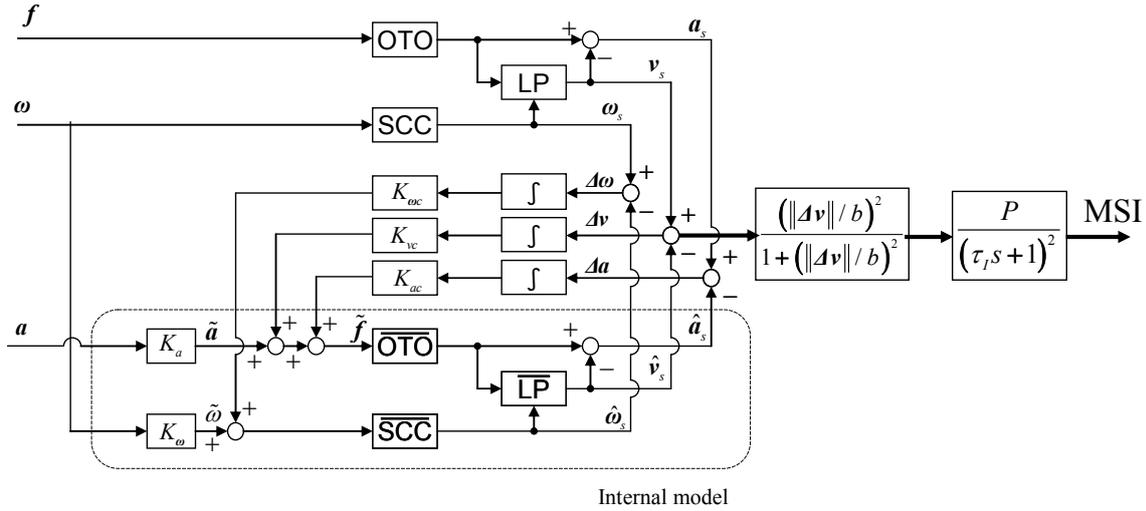

Figure 1: 6DOF-SVC model (modified from (Kamiji et al., 2007))

Drivers receive the acceleration stimulation and the rotational stimulation when negotiating a curve. The driver appropriately controls his/her posture of the head and the body by tilting his/her head to the direction of the curve turns. It should be noted that the head movement of the passenger is opposite to that of the driver. Moreover, because the driver does not get carsickness in comparison with the passenger, it is suggested that the driver's head movement is related to the decrease of carsickness and the improvement of ride comfort. Thus, we suppose that the mechanism of ride comfort can be understood by investigating the driver's active head tilt motion. In the present paper, the SVC model is applied to the investigation of the effect of head tilt on the motion sickness incidence.

In the present paper, the 6DOF-SVC model is first introduced to estimate motion sickness severity in 6-DOF motion in three-dimensional space (Kamiji et al, 2007). Second, by measuring the head tilt motions of passengers and drivers during slalom driving, the MSI of both drivers and passengers predicted by the 6DOF-SVC model are compared to investigate the effect of head tilt strategy on motion sickness.

## 2. 6DOF-SVC Model: Mathematical Model of Motion Sickness

The 6DOF-SVC model (Kamiji et al., 2007) is shown in Fig. 1. Please note that all vectors are seen from the head-fixed coordinate system. The inputs of the model are gravito-inertial acceleration (GIA) $f$ given in Eq. (1), angular velocity vector $\omega$, and inertial acceleration $a$.

$$f = a + g \qquad (1)$$

where $g$ denotes gravitational acceleration.

Vector $f$ is inputted to the otolith, denoted by the OTO block. The transfer function of the OTO is given by the unit matrix. Vector $\omega$ is inputted to the semicircular canals, denoted by the SCC block. The transfer function of the semicircular canal can be given by Eq. (2) (Haslwanter et al., 2000):

$$\omega_s^i = \frac{\tau_d \tau_a s^2}{(\tau_d s + 1)(\tau_a s + 1)} \omega^i \quad (i = x, y, z) \qquad (2)$$

where $\tau_a$ and $\tau_d$ are time constants.

The sensed vertical direction seen from the head-fixed frame $v_s$ is estimated from the otolith-canal interaction given by Eq. (3), denoted as LP in Fig. 1.

$$\frac{dv_s}{dt} = \frac{1}{\tau}(f - v_s) - \omega_s \times v_s \qquad (3)$$

where $v_s$ and $\omega_s$ denote the sensed vertical vector and the angular velocity vector of the head, respectively (Boss & Bless, 2002).

The lower part of the block diagram is the internal model of the vestibular system. Blocks $\overline{OTO}$ and $\overline{SCC}$ denote the internal model of OTO and SCC, respectively. The transfer function of $\overline{OTO}$ is also given by the unit matrix. The transfer function of $\overline{SCC}$ is given by Eq. (4) (Haslwanter et al., 2000).

$$\hat{\omega}_s^i = \frac{\tau_d s}{\tau_d s + 1} \tilde{\omega}_i \quad (i = x, y, z) \qquad (4)$$

where $\tau_d$ denotes the time constant that is the same as the constant used in Eq. (3).

It should be noted that the input of the internal model $\overline{OTO}$ is $a$, which is not measured from the vestibular system directly. Here we assume that a human roughly estimates body kinematics such as acceleration and angular velocity of the head from somatic sensation, including joint angles. Gain $K_a$ and $K_\omega$ represent estimation errors of the acceleration and the angular velocity, respectively.

The internal model of LP, illustrated as $\overline{\text{LP}}$, is assumed to be identical with LP. The outputs of the internal model are $\hat{a}_s$, $\hat{v}_s$, and $\hat{\omega}_s$. Vectors $\Delta a$, $\Delta v$, $\Delta \omega$ denote the error between the sensory information by the vestibular system ($a_s$) and the estimated information by the internal model ($\hat{a}_s$). These discrepancies are decreased through the adaptation process described by the integration with gains $K_\omega$, $K_v$, and $K_a$. Finally, the MSI is calculated based on the error between the sensed and estimated vertical direction $\Delta v$ through the Hill function and second-order lag with a large time constant. All parameters used in the model are given in (Kamiji et al., 2007).

The validity of the model is examined by comparing it with the reported experimental results (Donohew & Griffin, 2004), in which the moving base was vibrated for 2 hours at several frequencies. The results with the head rotation fit the experimental results better than the results without the head rotation, especially in low-frequency conditions. Please refer to Kamiji et al. (2007) and Wada et al. (2010) for more detail.

## 3. Predicted MSIs by Measured Head Movements of Drivers and Passengers

### 3.1. Experiments with a car

#### 3.1.1 Apparatus

An experiment with human subjects in a real car was conducted to measure driver and passenger head movements, and then the estimated MSI was calculated by the 6DOF-SVC model.

A small passenger car with a 2.46 m wheelbase and a 1300 cc engine was used for the driving experiments. A motion sensor (MTi-G, Xsense Technologies) was attached to a flat place close to the shift lever of the automatic transmission to measure the 3DOF acceleration and the 3DOF orientation of the vehicle. A gyro-type orientation sensor (Inertia Cube3, InterSense) and a wireless accelerometer (WAA-001, Wireless Technology) were attached to the cap worn by the participants to measure the 3DOF orientation and the 3DOF acceleration of the head. Both sensors were connected to a laptop PC in the rear seat of the vehicle to synchronize the sensor data. A straight asphalt road was used as the test track, in which five pylons were located at gaps of all 15 m or all 20 m.

#### 3.1.2 Design

Six males aged 22 to 24 yr who had a driver's license and who gave informed consent participated in the experiments as both drivers and passengers. The driver/passenger condition was treated as a within-subject factor. Each driver was asked to drive the pylon slalom on the straight road at a constant velocity. The following two conditions were set in the slalom driving:

Mild condition: Driving at 30 km/h with a 20 m gap between pylons
Hard condition: Driving at 40 km/h with a 15 m gap between pylons

#### 3.1.3 Procedure

Each participant was seated in the driving seat of the small passenger car in the normal driving position with a safety belt. The second participant was seated in the navigator seat in a normal seating position with a safety belt and then was exposed to the acceleration applied by the driver. The experimental course was a straight road with five pylons. The drivers were asked to drive in the predetermined constant velocity. Each driver drove the test track three times per pylon-velocity condition after some practice runs. Then, the participants drove at the other pylon-velocity condition three times after some practice runs. The order of the condition was randomized among the participants. The root mean square (RMS) vehicle lateral accelerations for the mild and hard conditions were 1.83 m/s$^2$ (SD 0.16) and 3.45 m/s$^2$ (SD 0.64), respectively.

### 3.2. Result

#### 3.2.1 Vehicle motion and head roll

Figs. 2 and 3 are examples of the lateral acceleration and the roll angle of the vehicle and those of the driver's and passenger's heads in the mild and hard conditions, respectively. It was found that the driver's head movement was in the opposite direction to the vehicle roll. The passenger's head roll was in the opposite direction to or with a large time lag to the driver's head roll in a passive manner, and not in good synchronization with the vehicle motion. The maximum head roll angles of the drivers and the passengers in the hard condition were larger than those in the mild condition.

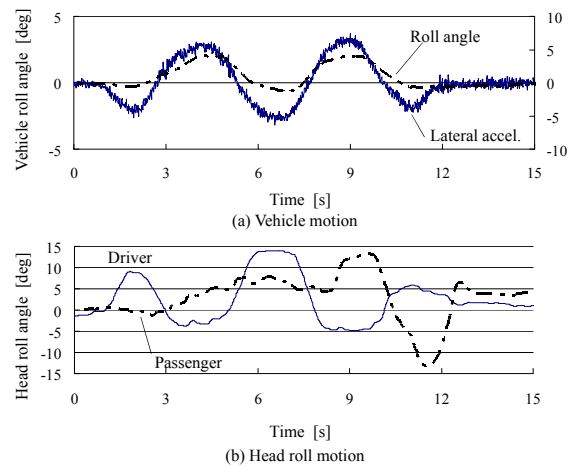

Figure 2: Vehicle motion and head motions (mild cond.)

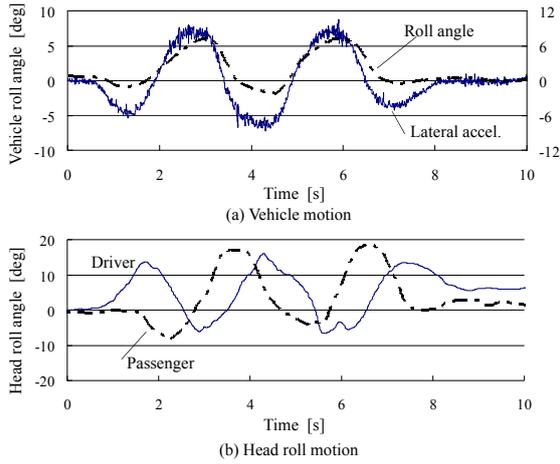

Figure 3: Vehicle motion and head motions (hard cond.)

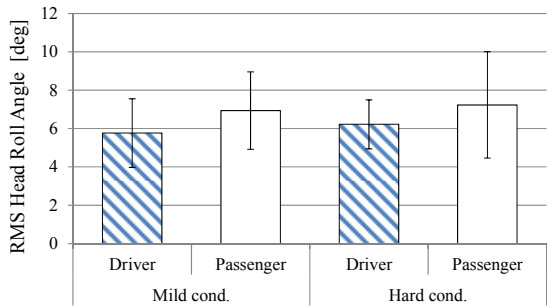

Figure 4: RMS head roll angles

Fig. 4 illustrates the RMS head roll angles. The Wilcoxon signed-rank test for each slalom condition revealed no significant differences between the driver and the passenger in both the mild and hard conditions. In addition, no significant differences were found in the head roll angle between the mild and hard conditions.

The correlation coefficients between the vehicle lateral acceleration and the head roll angle during the slalom driving were analyzed. Fig. 5 illustrates the mean correlation coefficients for all participants. The error bars represent the standard deviation (SD). The positive correlation coefficient indicates that the head roll motion synchronized well with the vehicle lateral acceleration in the centrifugal direction. A high negative correlation was found in the driver results, whereas a lower positive correlation was found in the passenger results. The Wilcoxon signed-rank test revealed the significance between the drivers and the passengers both in the mild condition ($z = 2.20$, $n = 6$, ties $= 0$, $p = 0.028$, two-tailed) and the hard condition ($z = 2.20$, $n = 6$, ties $= 0$, $p = 0.028$, two-tailed). The head roll angle of the drivers was greater in the centripetal direction, whereas that of the passengers was in the opposite direction at lower synchronization with the vehicle motion in the mild condition and at no synchronization in the hard condition (Fig. 5).

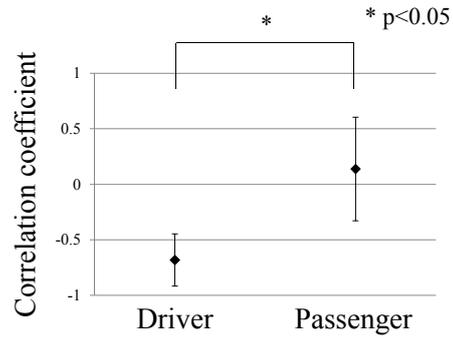

(a) Mild condition

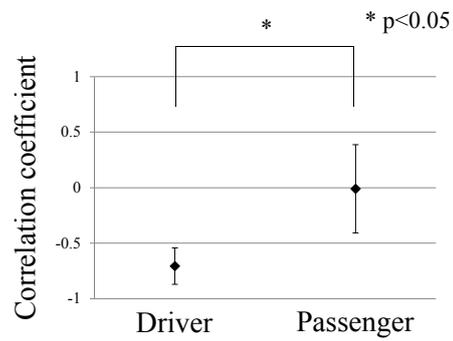

(b) Hard condition

Figure 5: Correlation coefficients between the vehicle lateral acceleration and the head roll angle

3.2.2 Predicted MSI with measured head motion

The predicted MSIs were calculated from the measured head motion in the experiments.

1) Driver-head condition: The angular velocity of the driver's head, as calculated from the experimental results, was inputted to the 6DOF-SVC model.
2) Passenger-head condition: The angular velocity of the passenger's head, as calculated from the experimental results, was inputted to the 6DOF-SVC model.

The acceleration of the vehicle was used as the input of the head acceleration for all conditions due to the large amount of noise present in the driver's and passenger's head acceleration. The MSIs were calculated by using the experimental results during slalom driving.

Fig. 6 shows the results of the predicted MSI for each condition. The Wilcoxon signed-rank test revealed the significance between the drivers and the passengers in the mild condition ($z = 1.99$, $n = 6$, ties $= 0$, $p = 0.046$, two-tailed). The predicted MSI calculated from the drivers' head movement in the experiments was smaller than that calculated from the passengers. In contrast, no significance was found in the hard condition ($z = 1.36$, $n = 6$, ties = 0, $p = 0.173$, two-tailed).

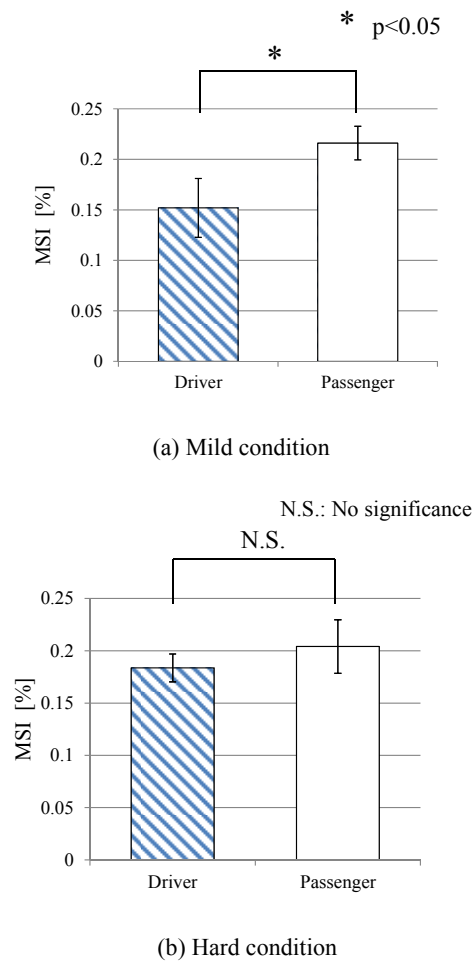

(a) Mild condition

(b) Hard condition

Figure 6: Predicted MSIs

## 4. Discussion

The experimental results with a real car demonstrated that the drivers tilted their head toward the centripetal direction, which is the opposite tendency of the passengers. The findings of the present paper are consistent with the observation of the head movements of bus drivers and passengers by Fukuda (1976) as well as the experimental results with real car experiments by Zikovitz and Harris (1999).

Then, the predicted MSIs were calculated by the 6DOF-SVC model by inputting the experimental results. The results revealed that the MSIs of the drivers were significantly smaller than those of the passengers. These findings agree well with the estimated MSI in the supposed head motion generated from the simulated vehicle motion under slalom driving (Wada et al., 2010). In addition, they agree with the results from Golding et al. (2003), in which the severity of motion sickness decreased when the active head movement of the participant was aligned with the GIF in a longitudinal linear acceleration environment.

The contribution of the present paper is that it demonstrates the effect of head tilt on reducing the severity of the MSI in a real automotive environment under lateral acceleration and by using the 6DOF-SVC model. No significance was found in the hard condition. It can be understood that it was difficult for the drivers to tilt their head appropriately to reduce the severity of the MSI in the hard condition. This result suggests that the severity of the MSI of the passengers can be reduced by tilting their head toward the centripetal direction, as was done by the drivers. An experimental study with a real car, conducted by Wada et al. (2012), agrees with this idea. That study also revealed that the severity of the motion sickness was reduced significantly when the passenger imitated the driver's head movement during slalom driving.

It should be noted that the 6DOF-SVC model does not include visual and somatosensory information, because it is limited to the vestibular sensory system.

## 5. Conclusion

The 6DOF-SVC model was introduced to estimate the MSI in 6-DOF motion and it was applied to the investigation of the head tilt strategy on the motion sickness incidence in real car experiments. The results revealed that the estimated MSI of the driver was smaller than that of the passenger. It was suggested that the driver's head movement toward the centripetal direction has the effect of reducing motion sickness. This result gives us a new interpretation of the driver's head movements; that is, the head movement reduces the MSI in the sense of SVC theory. This result suggests that the severity of the MSI of the passengers can be reduced by the passengers tilting their head toward the centripetal direction, as is done by the drivers.

## References

Bless, W., Boss, J. E., De Graf, B., Groan, E., and Wertheim, H. (1998). Motion sickness: Only one provocative conflict? *Brain Research Bulletin*, 47(5), 481–487.

Bos, J. E., and Bles, W. (1998). Modeling motion sickness and subjective vertical mismatch detailed for vertical motions. *Brain Research Bulletin,* 47(5), 537–542.

Bos, J. E., and Bles, W. (2002). Theoretical considerations on canal-otolith interaction and an observer model. *Biological Cybernetics*, 86, 191–207.

Donohew, B. E., and Griffin, M. J. (2004). Motion sickness: Effect of the frequency of lateral oscillation. *Aviation Space and Environmental Medicine*, 75(8), 649–656.

Fukuda, T. (1976). Postural Behaviour and Motion Sickness, *Act Oto-laryngologica*, 81, 237–241.

Golding, J. F., Bles, W., Bos, J. E., Haynes, T., and Gresty, M.A. (2003). Motion sickness and tilts of the inertial force environment: Active suspension system vs. active passengers, *Aviation, Space, and Environmental Medicine*, 74(3), 220–227.

Griffin, M. J. (1990). *Handbook of Human Vibration*. London: Academic Press.

Haslwanter, T., Jaeger R., Mayr, S., and Fetter, M., (2000). Three dimensional eye movement responses to off-vertical axis rotations in human. *Experimental Brain Research.* 134(1), 96–106.

Kamiji, N., Kurata, Y., Wada, T., and Doi, S. (2007). Modeling and validation of carsickness mechanism. *Proceedings of International Conference on Instrumentation, Control and Information Technology*, 1138–1143.

Muragishi, Y., Fukui, K., Ono, E., Kodaira, T., Yamamoto, Y., and Sakai, H. (2007). Improvement of vehicle dynamics based on human sensitivity (First Report) -Development of human sensitivity evaluation system-. *SAE paper 2007-01-0448, Proceedings of SAE2007 World Congress*.

O'Hanlon, J. F., and McCauley, M. E. (1974). Motion sickness incidence as a function of the frequency and acceleration of vertical sinusoidal motion. *Aerospace Medicine*, 5(4), 366–369.

Reason, J. T., Brand, J. J. (1975). *Motion Sickness*. Academic Press, London.

Wada, T., Fujisawa, S., Imaizumi, K., Kamiji, N., and Doi, S. (2010). Effect of driver's head tilt strategy on motion sickness incidence, *Proceedings of 11th IFAC/IFIP/IFORS/IEA Symposium on Analysis, Design, and Evaluation of Human-Machine Systems*, CD-ROM (6 pages).

Wada, T., Konno, H., Fujisawa, S., and Doi, S. (2012) Can passenger's active head tilt decrease the severity of carsickness?: Effect of head tilt on severity of motion sickness in a lateral acceleration environment, *Human Factors*, 54(2), 226–234.

Zikovitz, D. C., and Harris, L. R. (1999). Head tilt during driving. *Ergonomics*, 42(5), pp. 740–746.